\documentclass[aps,prd,twocolumn,superscriptaddress,amsmath,amssymb]{revtex4}

\usepackage{graphicx} % Include figure files
\usepackage{epstopdf} % Allow to include eps files
\usepackage{dcolumn} % Align table columns on decimal point
\usepackage{xcolor} % Color support
\usepackage{amsmath,amssymb} % Extended symbol collection
\PassOptionsToPackage{hyphens}{url}\usepackage[hidelinks,breaklinks=true]{hyperref} % Generate pdfs with hyperlinks
\usepackage[utf8]{inputenc} % Allow UTF8 characters
\usepackage[english]{babel} % All publications are in English
\usepackage[displaymath, mathlines]{lineno} % Line numbers for drafts
\usepackage{blindtext} % For testing
\usepackage{ifthen} % for conditional statements
\usepackage{orcidlink} % for ORCIDs in author list
\usepackage{multirow}
\usepackage{gensymb}
\newboolean{articletitles}
\setboolean{articletitles}{true} % False removes titles in references

\usepackage{environ}
\NewEnviron{comment}{}

\newcommand\coff{\RenewEnviron{comment}{\BODY}}

\coff

\newcommand{\changecolor}{red}

\newcommand{\changeoff}{\renewcommand{\changecolor}{black}}

\newcommand{\changecolortwo}{blue}

\newcommand{\changetwooff}{\renewcommand{\changecolortwo}{black}}

\newcommand{\changecolorthree}{blue}

\newcommand{\changethreeoff}{\renewcommand{\changecolorthree}{black}}

\changeoff
\changetwooff

\changethreeoff
\input{belle2-symbols}

\begin{document}

% \pacs{\input{pacs}}
\title{Observation of the decay \jpsiomega at \belletwo}

% Leo will provide the list of authors during CWR
%%% Paper:    B0 to J/psi omega
%%% Journal:  Physical Review D
%%% Contacts: N. Brenny, C. Chen, M. Veronesi
%%% ====================================================================
%%% Use \input{pub071-orcid} to insert this material into your latex file.
  \author{I.~Adachi\,\orcidlink{0000-0003-2287-0173}} % 2590
% \author{K.~Adamczyk\,\orcidlink{0000-0001-6208-0876}} % 2239
  \author{L.~Aggarwal\,\orcidlink{0000-0002-0909-7537}} % 10083
% \author{P.~Ahlburg\,\orcidlink{0000-0002-9832-7604}} % 2367
  \author{H.~Ahmed\,\orcidlink{0000-0003-3976-7498}} % 11323
% \author{J.~K.~Ahn\,\orcidlink{0000-0002-5795-2243}} % 7423
  \author{H.~Aihara\,\orcidlink{0000-0002-1907-5964}} % 2223
  \author{N.~Akopov\,\orcidlink{0000-0002-4425-2096}} % 9443
  \author{M.~Alhakami\,\orcidlink{0000-0002-2234-8628}} % 28103
  \author{A.~Aloisio\,\orcidlink{0000-0002-3883-6693}} % 2194
  \author{N.~Althubiti\,\orcidlink{0000-0003-1513-0409}} % 21524
% \author{L.~Andricek\,\orcidlink{0000-0003-1755-4475}} % 2098
% \author{M.~Angelsmark\,\orcidlink{0000-0003-4745-1020}} % 13963
  \author{N.~Anh~Ky\,\orcidlink{0000-0003-0471-197X}} % 2218
  \author{D.~M.~Asner\,\orcidlink{0000-0002-1586-5790}} % 4684
% ...
% 13403
% \author{S.~Zou\,\orcidlink{0000-0003-3377-7222}} % 19363
\collaboration{The Belle II Collaboration}

\begin{abstract}
We measure the branching fraction of the decay \jpsiomega using data collected with the \belletwo detector at the SuperKEKB collider.
The data contain \nbb \BBbar meson pairs produced in energy-asymmetric \epem collisions at the \FourS resonance.
The measured branching fraction $\bf(\jpsiomega)=\left( \bfval \pm \bferr \pm \bfsys \right) \times 10^{\bfpower}$, where the first uncertainty is statistical and the second is systematic, is more precise than previous results and constitutes the first observation of the decay with a significance of $\sensitivitysysnosigma$ standard deviations.
\end{abstract}

\maketitle

\section{Introduction}
Decays of \B mesons into a charmonium state and a light unflavored meson are predominantly governed by color-suppressed tree diagrams involving the quark transition ${\bquark \to \Pc \overline \Pc \dquark}$. 
One such decay, \jpsiomega, has not been observed yet.
This mode can be used as a control channel in studies of \B decays mediated by the ${\bquark \to \dquark \ellell}$ transition at a \B-factory.
Since the quark-level transition of this decay is the same as that of ${\Bz \to \jpsi \piz}$, its measured \CP asymmetries can be used in a similar way to constrain the contributions from loop diagrams in ${\Bz \to \jpsi \Kz}$, as in Refs.~\cite{PhysRevLett.95.221804,Barel_2021}.

The \lhcb experiment reported the first evidence for \jpsiomega by reconstructing the mode ${\omega \to \pip \pim \piz}$~\cite{LHCb:2012cw}.
They found a significance of 4.6 standard deviations ($\sigma$) using the negative log likelihood scan and taking into account systematic uncertainties related to the fit function.
Furthermore, \lhcb studied the resonant structure of the decay ${\Bz \to \jpsi \pip \pim}$ using an amplitude analysis and found the branching fraction to be $\BF(\jpsiomega)=(1.8^{+0.7}_{-0.5})\times 10^{-5}$ with the $\omega$ decaying to $\pip \pim$~\cite{PhysRevD.90.012003}.

We present a measurement of the branching fraction of the decay \jpsiomega using a sample of energy-asymmetric \epem collisions at the \FourS resonance provided by the SuperKEKB accelerator~\cite{Akai:2018mbz} and collected with the \belletwo detector~\cite{Abe:2010gxa}.
This sample has an integrated luminosity of \lumi and contains \nbb \BBbar events~\cite{thebelleiicollaboration2024measurementintegratedluminositydata}.
We also use an off-resonance sample totaling \offreslumi recorded 60 \mev below the \FourS resonance to study the background from the continuum ${\epem \to \qqbar}$ events, where \quark denotes a \uquark, \dquark, \squark, or \cquark quark. 

We reconstruct \Bz mesons in the $\jpsi \omega$ final state using the subdecays \jpsill (with $\ell$ being an electron or a muon), ${\omega\to\pip\pim\piz}$, and ${\piz \to \gamma \gamma}$.
We extract the signal yields by fitting distributions of signal candidates in a kinematic observable that discriminates against backgrounds.
We validate our analysis procedure on simulated samples and correct for differences between collision data and simulation using control samples.
Charge-conjugated modes are included throughout the paper unless explicitly stated.
\section{Experiment}
The \belletwo detector operates at the SuperKEKB accelerator at KEK, which collides 7~\gev electrons with 4~\gev positrons.
The detector is designed to reconstruct the decay products of heavy-flavor hadrons and $\tau$ leptons.
It consists of several subsystems arranged cylindrically around the interaction point (IP).
The innermost part of the detector is equipped with a two-layer silicon-pixel detector (PXD), surrounded by a four-layer double-sided silicon-strip detector (SVD)~\cite{Belle-IISVD:2022upf}.
Together, they provide information about charged-particle trajectories (tracks) and decay-vertex positions.
Of the outer PXD layer, only one-sixth is installed for the data used in this work.
The momenta and electric charges of charged particles are determined with a 56-layer central drift-chamber (CDC).
Charged-particle identification (PID) is provided by a time-of-propagation counter and an aerogel ring-imaging Cherenkov counter, located outside the CDC in the barrel and forward regions, respectively.
The CDC provides additional PID information through the measurement of specific ionization.
Photons are identified and electrons are reconstructed by an electromagnetic calorimeter (ECL) made of CsI(Tl) crystals, covering the region outside of the PID detectors.
The tracking and PID subsystems, and the calorimeter, are surrounded by a superconducting solenoid, providing an axial magnetic field of 1.5~T.
The central axis of the solenoid defines the $z$ axis of the laboratory frame, pointing approximately in the direction of the electron beam.
Outside of the magnet lies the muon and \KL identification system, which consists of iron plates interspersed with resistive-plate chambers and plastic scintillators.

We use Monte Carlo (MC) simulated events to model signal and background distributions, study the detector response, and test the analysis procedure.
Quark-antiquark pairs from \epem collisions are generated using \kkmc~\cite{Jadach:1999vf} with \pythia~\cite{Sjostrand:2014zea}.
Signal and other \B-meson decays are generated with \evtgen~\cite{Lange:2001uf}.
Effects of final-state radiation are incorporated with \photos~\cite{Barberio:1990ms}.
The detector response is simulated with \geant~\cite{Agostinelli:2002hh}.
We use a simulated sample of generic \epem collisions, corresponding to an integrated luminosity of approximately four times that of the experimental dataset.
We also use large samples of simulated \BBbar pairs in which one of the \B mesons is forced to decay to the final state of interest, while the other \B meson in the event decays inclusively.
One sample is used to study the signal, where the \B meson decays as \jpsiomega.
The other samples are used to study the dominant source of background, where the \B meson decays inclusively into ${\B \to \jpsi X}$ modes.
We process collision data and simulated samples using the \belletwo analysis software~\cite{Kuhr:2018lps,basf2-zenodo}.
\section{Event Selection}
Events containing a \BBbar pair are selected by a hardware trigger system based on the track multiplicity and total energy deposited in the ECL.
The trigger efficiency is close to 100\% for signal decays.

We reconstruct $\jpsiomega$ candidates using the decay chain ${\jpsi\to\ellell}$, ${\omega\to\pip\pim\piz}$, and ${\piz\to\gamma\gamma}$.
Tracks are reconstructed using information from the PXD, SVD, and CDC~\cite{Bertacchi:2020eez}. 
Each track must have a polar angle $\theta$ within the CDC acceptance ($17^\circ<\theta<150^\circ$) and is also required to have a distance of closest approach to the IP less than $2.0$~\cm along the $z$ axis and less than $0.5$~\cm in the transverse plane to reduce contamination from misreconstructed and beam background-induced tracks.
Electron identification is provided by a boosted decision tree (BDT) classifier that combines several ECL variables and PID likelihoods~\cite{ebdt}.
Muons are identified with the discriminator $\prob_{\mu} = \lh_{\mu}/(\lh_{e}+\lh_{\mu}+\lh_{\pi}+\lh_{K}+\lh_{d}+\lh_{p})$, where the likelihood $\lh_{i}$ for each charged particle hypothesis combines PID information from all subdetectors except for the SVD and PXD.
We classify tracks as electrons or muons based on a PID requirement that is $\eideff$ $(\muideff)$ efficient on signal while rejecting $\eidrej$ $(\muidrej)$ of tracks for electrons (muons).
The momenta of electrons are corrected for energy loss due to bremsstrahlung by adding the four-momenta of photons detected within 50~\mrad of the initial direction of the electron tracks.
Charged pions are selected by requiring the pion likelihood ratio $\lh_{\pi}/(\lh_{\pi}+\lh_{K})$ to be greater than a certain threshold, which retains $\hadideff$ of signal pions while rejecting $\hadidrej$ of kaons.

The \jpsi candidates are formed by combining oppositely charged lepton pairs having an invariant mass $\mee\in[2.95,3.15]$~\gevcc and $\mmumu\in[3.0,3.15]$~\gevcc, where the average \jpsi invariant mass resolution is $16$ ($13$)~$\mevcc$ in the \epem (\mumu) mode.

Photons are identified from ECL energy deposits greater than $80$, $40$ and $30$~\mev in the forward, backward and barrel regions, respectively.
The corresponding polar-angle coverages are [12.4, 31.4]$\degree$,  [130.7, 155.1]$\degree$, and [32.2, 128.7]$\degree$.
The ECL energy deposits are required to have no matched tracks in the CDC.
Photon energy corrections are derived from an ${\epem \to \mumu \gamma}$ control sample reconstructed in data.
The \piz candidates are formed by combining pairs of photons with an invariant mass $\mgg\in[0.120,0.145]$~\gevcc.
The average \piz invariant mass resolution is 8~\mevcc.

The $\omega$ candidates are formed by combining \piz candidates with two oppositely charged pions with an invariant mass $m(\pip\pim\piz)\in[0.73,0.83]$~\gevcc.
The average $\omega$ invariant mass resolution is 15~\mevcc.

The \jpsiomega decay vertex is determined using the \treefit algorithm~\cite{HULSBERGEN2005566, Krohn:2019dlq}.
The \jpsi invariant mass is constrained to its known value~\cite{ParticleDataGroup:2024cfk}.
We retain \B candidates with a successful vertex fit.
The beam-energy constrained mass \mbc and energy difference \deltae are computed for each \jpsiomega candidate as $\mbc \equiv \sqrt{(\ebeam/c^2)^2-(|\pstar|/c)^2}$ and $\deltae \equiv \estar - \ebeam$, where \ebeam is the beam energy, and \estar and \pstar are the energy and momentum of the \Bz candidate, respectively, all calculated in the center-of-mass (c.m.)\ frame.
Signal \Bz candidates peak at the known \Bz mass~\cite{ParticleDataGroup:2024cfk} in \mbc and zero in \deltae. 
The average \mbc and \deltae resolutions for correctly reconstructed signal events are 3~\mevcc and 20~\mev, respectively. 
Candidates satisfying $\mbc>5.27~\gevcc$ and $|\deltae|<0.25~\gev$ are retained for further analysis.

We suppress the contribution from the dominant source of background, $\jpsix$ decays that do not contain a true $\omega$ meson, using BDT classifiers~\cite{10.1145/2939672.2939785} that combine several variables to discriminate between signal and background. 
We use a separate BDT for each \jpsi decay mode because the background shapes are mode-dependent.
The BDTs are trained on the following six input variables: the momentum of the \piz candidate in the laboratory frame, the invariant mass of the $\omega$ candidate, the invariant masses of the $\piz\pip$ and $\piz\pim$ systems, and the invariant masses of the $\pip\pim$ system, calculated by assigning a kaon mass hypothesis to the positively (negatively) charged pion while retaining a pion mass hypothesis for the negatively (positively) charged pion. 

To validate the MC modeling of BDT input variables, we compare their distributions in the \mbc and \deltae sidebands in simulation and data and find them to agree well. 
The \mbc sidebands correspond to the region $5.20 < \mbc < 5.25~\gevcc$, which excludes 100\% of the simulated signal.
The \deltae sidebands are chosen so that the central 99.97\% of the simulated signal is excluded: $\deltae \in [-0.250,-0.148]\cup[0.102,0.250]~\gev$ in the \epem mode and $[-0.250,-0.147]\cup[0.094,0.250]~\gev$ in the \mumu mode. 
We train the BDTs using signal and background samples from simulated data. 
For each BDT, we choose a criterion to be applied on its output by optimizing a figure of merit $\rm S/\sqrt{\rm S + \rm B}$, where $\rm S$ and $\rm B$ are the number of correctly-reconstructed signal and background candidates in the \deltae region containing the central 95\% of the signal.
In the calculation of the figure of merit we use the current world-average values of the signal and ${\Bz \to \jpsi X}$ branching fractions~\cite{ParticleDataGroup:2024cfk}.
The BDT selection retains $\bdteffee$ ($\bdteffmumu$) of the signal while rejecting $\bdtrejee$ ($\bdtrejmumu$) of the background in the \epem (\mumu) mode.

Events with more than one candidate account for approximately $9\%$ of the events selected in the collision data.
For events with multiple candidates, we retain the one with the highest BDT score.
This requirement selects the correct signal candidate $68\%$ of the time for events with multiple candidates in simulation.

To account for data-MC differences in the signal probability density function, we reconstruct a control mode with the same final state, ${\Bz \to \jpsi \Kstarz}$, with ${\Kstar \to \KS \piz}$, ${\KS \to \pip \pim}$, and ${\piz \to \gamma\gamma}$.
We apply the same event selection, except for the requirements on the pion tracks from the \KS, since the \KS reconstruction is cleaner than the $\omega$ reconstruction.
We also use a different $m(\pip\pim\piz)$ requirement.
The \KS candidates are formed by combining oppositely charged tracks and requiring $m(\pip\pim)\in[0.45,0.55]$~\gevcc.
The \Kstarz candidates are required to have an invariant mass $m(\pip\pim\piz)\in[0.817,0.967]~\gevcc$.
We also require the cosine of the angle between the \Bz and $\piz$ momenta in the \Kstarz rest frame to be greater than $-0.7$.

To account for data-MC differences in the efficiency of the \jpsi vertex fit, we reconstruct an inclusive \jpsi control mode by applying the same \jpsi selection as in the signal mode.

To validate the BDT, we reconstruct another control mode, ${\Dz \to \KS \omega}$, with ${\omega \to \pip \pim \piz}$ and ${\KS \to \pip \pim}$.
Since the BDT input variables depend only on the decay products of the $\omega$, and all but the laboratory-frame \piz momentum are Lorentz invariant, we choose a control mode that has high $\omega$ purity in order to calibrate the signal efficiency of the BDT selection.
We apply the same $\omega$ selection as in the signal mode and the same \KS selection as in the ${\Bz \to \jpsi \Kstarz}$ control mode. 
The \Dz candidate is formed by combining the $\omega$ and \KS using the \treefit algorithm~\cite{HULSBERGEN2005566, Krohn:2019dlq}.
We require the $\chi^2$ probability of the \Dz vertex fit to be greater than 0.1\%.
A \Dstarp candidate is formed by combining a track with a pion mass hypothesis and the \Dz candidate.
We suppress background from \BBbar decays by requiring the momentum of the \Dstarp candidates in the \FourS c.m.\ frame to be greater than 2.4 \gevc.
We further require that the mass difference between the \Dstarp and \Dz candidates be in the range [0.1445,0.1465]~\gevcc.
The BDT input variables' distributions in reconstructed ${\Dz \to \KS \omega}$ decays show good agreement between data and simulation.

The signal efficiencies corrected for differences between data and simulation are listed in Table~\ref{tab:de-fit}.
\section{Signal Extraction Fit}
The sample passing the event selection is populated by \jpsiomega candidates coming from both signal and backgrounds.
For signal events with multiple candidates, simulation studies show that incorrectly selected signal candidates, based on their BDT scores, are predominantly due to misreconstructed $\omega$ mesons with a low-momentum ${\piz \to \gamma \gamma}$ originating from the other \B meson decay.
The \deltae distribution of these incorrectly selected \Bz candidates is centered around zero and resembles the distribution of correctly reconstructed signal candidates.
Consequently, they are also treated as signal candidates in the measurement.

Among various sources of background, the largest contribution comes from ${\B \to \jpsi X}$ decays, for which the \jpsi is correctly reconstructed but originates from a different \B decay than the signal.
There is also a contribution from \BBbar events with misreconstructed \jpsi and continuum backgrounds, which accounts for 4\% (15\%) of candidates in the \epem (\mumu) mode in simulation.
The \mumu mode has a higher background here since the muon PID requirement has a higher fake rate than that of the electron.
We validate our suppression of continuum background by applying the selection criteria to the off-resonance sample. We find a background yield consistent with our expectation from simulation.

The selection efficiency in simulation for nonresonant ${\Bz\to\jpsi\pip\pim\piz}$ decays generated with a phase space model is less than 1\% of our signal selection efficiency. 
The branching fraction of these nonresonant decays has not been measured. 
From a fit to the \deltae distribution in the $\omega$ invariant mass sidebands, $m(\pip\pim\piz)\in[0.60,0.73] \cup [0.83,1.00]$ \gevcc, we expect a contribution from nonresonant decays of $\nonresyieldee\pm\nonresyielduncee$ ($\nonresyieldmm\pm\nonresyielduncmm$) in the $\omega$ invariant mass signal window in the \epem (\mumu) mode.
We correct the fitted signal yields by the central values and take the statistical uncertainties as a systematic uncertainty.
The interference between the signal and nonresonant decays is calculated assuming the latter branching fraction to be the same as that of the signal and its impact on the signal yields is found to be negligible.

We extract the signal yields from an extended maximum-likelihood fit to the unbinned \deltae distributions.
The probability density function (\pdf) of the signal is described by a double-sided Crystal Ball ($F_{\rm CB}$) function~\cite{Gaiser:Phd,Skwarnicki:1986xj} with all parameters excluding the mean $\mu$ determined from simulation: $\sigma$, $\alpha_L$, $n_L$, $\alpha_R$, and $n_R$.
We account for the data-MC differences by scaling the width $\sigma$ of the Gaussian core of $F_{\rm CB}$ with a scale factor obtained by fitting the ${\Bz \to \jpsi \Kstarz}$ control sample. 
The difference in \deltae shape between the two \jpsi decay modes is negligible.
Therefore, the parameters of the signal \pdf are shared between \jpsiee and \jpsimm modes.
Each background component has a smooth \deltae distribution, enabling us to model the distribution of the sum of all backgrounds using an exponential \pdf ($F_{\rm exp}$).

\begin{figure}[!htb]
\centering
\includegraphics[width=0.45\textwidth]{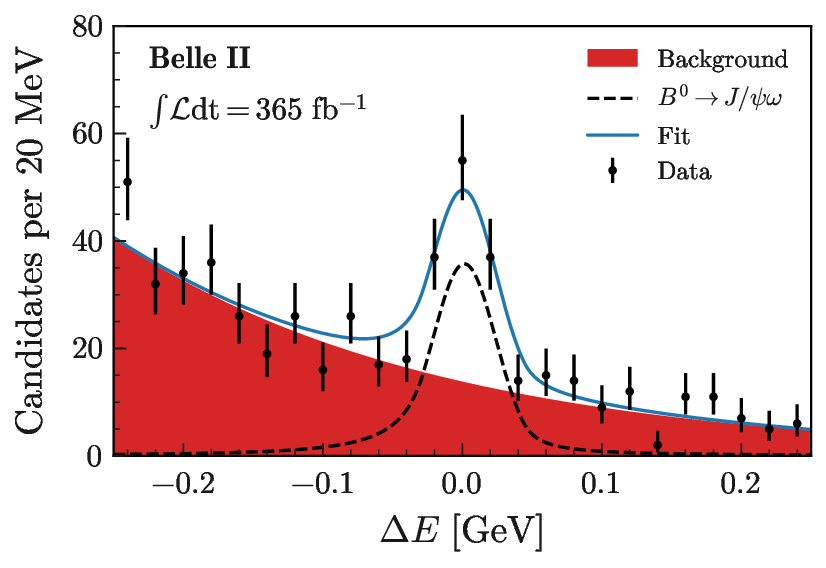}
\caption{Distribution of \deltae for \jpsiomega candidates (data points) with fit overlaid (curves and filled area). The background component includes contributions from ${\Bz\to\jpsi X}$, other \BBbar decays, and continuum events.}
\label{fit:de-fit}
\end{figure}

In the fit to the data, we determine the signal and background yields, \nsig and \nbkg, separately for the \jpsiee and \jpsimm modes, the mean $\mu$ of the signal \pdf, and the exponential decay parameters of the background \pdf, $c^i$.
In total, the fit has seven free and five fixed parameters.
The likelihood function is

\begin{equation}
\label{eq:pdf}
        \mathcal{L} = \frac{e^{-(\nsig+\nbkg)}}{N!} \prod_{i=1}^{N} \{ \nsig F^{i}_{\rm CB} + \nbkg F^{i}_{\rm exp} \}
\end{equation}

where $i$ is the index of the candidate, $N$ is the total number of candidates in the dataset, and $F^{i}_{\rm CB}$ and $F^{i}_{\rm exp}$ are the signal and background \pdfs of the $i$th candidate, respectively.  
In total, we fit 262 (274) candidates in the \epem (\mumu) mode.
The data and the fit are shown in Fig.~\ref{fit:de-fit}.
Table~\ref{tab:de-fit} lists the signal selection efficiencies, fitted yields of signal (corrected for nonresonant contribution) and background, as well as the statistical significances that are calculated by dividing the central values by the statistical errors.

From the signal yields, we determine the branching fraction:
\begin{equation}
\BF = \frac{\left( \nsigeeprime/\effee + \nsigmumuprime/\effmumu \right) (1+\fpm/\fzz)}{\BF(\jpsi \to \ellell)\BF(\omega \to \pip \pim \piz)\BF(\piz \to \gamma \gamma)2\numbb}
\end{equation}
where $\eff_{\text{sig}}$ are the efficiencies obtained from simulated signal samples and corrected for data-MC differences using control samples, ${\BF(\jpsi \to \ellell)}$ is the sum of $\BF(\jpsiee)=(5.971\pm0.032)\%$ and $\BF(\jpsimm)=(5.961\pm0.033)\%$, ${\BF(\omega\to\pip\pim\piz)}$ is $(89.2\pm 0.7)\%$, ${\BF(\piz\to\gamma\gamma)}$ is $(98.823\pm0.034)\%$~\cite{ParticleDataGroup:2024cfk}, $\numbb=\nbb$ is the number of \BBbar pairs in the dataset, and $\fpm/\fzz=1.052\pm0.031$~\cite{banerjee2024averagesbhadronchadrontaulepton} is the ratio of branching fractions for the \FourS decays to $\PB^+\PB^-$ and $\B^0\Bzb$.
We obtain $\BF(\jpsiomega) = (\bfval \pm \bferr) \times 10^{\bfpower}$, where the uncertainty is statistical only.

\begin{table}[!htb]
\caption{Signal efficiencies corrected for data-MC differences (uncertainties are systematic), signal yields corrected for nonresonant contribution and background yields (uncertainties are statistical), as well as statistical significances.}
\label{tab:de-fit}
\begin{tabular*}{\linewidth}{@{\extracolsep{\fill}}lcccc}
\hline
\hline
Decay mode  & $\eff_{\text{sig}}$ [\%] & $n'_{\text{sig}}$ & $n_{\text{bkg}}$ & stat. sig. [$\sigma$]\\
\hline
\jpsiee & $\eeeff$ & $\eeyield$ & $\eebkgyield$ & \eesignificancenosigma\\
\jpsimm &$\mumueff$ & $\mumuyield$ & $\mumubkgyield$ & \mumusignificancenosigma\\
\hline
\hline
\end{tabular*}
\end{table}
\section{Systematic Uncertainties}
Table~\ref{tab:bf-sys} lists contributions from various sources of systematic uncertainty.
Below, we describe each of them as well as the procedure to estimate their contributions.

In the calculation of the branching fraction, we correct the signal efficiencies obtained in simulation using control samples in the data.
The statistical and systematic uncertainties associated with the correction factors are propagated together as a systematic uncertainty on the measured branching fraction.

Tracking efficiencies are measured with ${\epem\to \taup\taum}$ events, where one $\tau$ decays as ${\taum \to e^- \overline{\nu}_e\nu_{\tau}}$ and the other as ${\taum \to \pim\pip\pim \nu_{\tau}}$.
A difference of $0.27\%$ for each track is observed between data and simulation, which is propagated to the uncertainty on the branching fraction.

The \piz reconstruction efficiency is measured in data and simulation from the ratio of the yields of ${\Dstarp\to \Dz(\to \Km \pip \piz)\pip}$ and ${\Dstarp\to \Dz(\to \Km \pip)\pip}$, scaled by the inverse of their branching fractions.
The yield ratio in collision and simulated data is used to obtain correction factors as a function of the $\piz$ momentum and polar angle.
The average correction factor of the \piz efficiency in \jpsiomega decays is \pizcorr, with the uncertainty dominated by the knowledge of the \Dz-decay branching fractions~\cite{ParticleDataGroup:2024cfk}.

The data-MC difference in electron and muon identification performance is calibrated with ${\jpsi \to \ell^+\ell^-}$, ${\epem \to \ell^+\ell^-(\gamma)}$, and ${\epem\to \epem \ell^+\ell^-}$ samples.
The average correction factor over the kinematic distribution of the signal is $\eidcorr$ ($\midcorr$) for the \epem (\mumu) mode, where the uncertainties are calculated as the quadrature sum of the statistical and systematic uncertainties.

The difference in pion identification performance between simulation and data is calibrated with samples of ${\Dz\to \Km \pip}$ and ${\KS\to\pip\pim}$ decays.
The uncertainty on the average correction factor over the kinematic distribution of the signal is negligible.

The vertex fit efficiencies for the \jpsi and $\omega$ are determined by calculating the ratio of signal yields with and without the vertex constraint for both collision and simulated events using inclusive $\jpsi$ and ${\Dz \to \KS \omega}$ control samples.
We fit the invariant mass distributions of the selected $J/\psi$ and $\omega$ candidates from the control samples to extract the signal yields.
The measured efficiencies in data and simulation are consistent within statistical uncertainties. 
The statistical uncertainties on the data-MC ratio of efficiencies are assigned as the systematic uncertainties due to vertex fit constraints.
Additionally, we measure the efficiency of the $\omega$ invariant mass window by fitting the corresponding distribution from the ${\Dz \to \KS \omega}$ control sample.
The efficiencies obtained in data and simulation are statistically consistent, and the statistical uncertainty on their ratio is taken as the systematic uncertainty associated with the $\omega$ invariant mass window selection.

The performance of the fake $\omega$ suppression BDT is validated with the control mode ${\Dz\to \KS\omega}$.
The ratio of the signal efficiency in data and simulation due to the BDT requirement is $\bdteecorr$ ($\bdtmmcorr$) in the \epem (\mumu) mode, where the uncertainty is statistical only.
The uncertainty, including the correlation between corrections for the two \jpsi decay modes, is propagated to the uncertainty on the branching fraction.

The signal efficiency depends on the longitudinal polarization fraction of the signal decay.
We simulate the signal decay with \evtgen using helicity amplitudes measured in Ref.~\cite{PhysRevD.90.012003}. 
We take the difference in the signal efficiency when varying the helicity amplitudes around their uncertainties as a systematic uncertainty.

We consider the following uncertainties associated with the signal yields obtained from the fit. 
We assign a systematic uncertainty due to imperfect signal modeling using simplified simulated datasets in which the signal is sampled from simulated data and the background is generated with the exponential \pdf. 
The systematic uncertainty is assigned to be the product of the statistical uncertainty from the fit and the mean of the distribution of the difference of the fit yield from its true value divided by the fit error. 
To assess the systematic uncertainty associated with fixing the width parameter in the signal \pdf, we repeat the fit by varying the scale factor determined in the ${\Bz \to\jpsi\Kstarz}$ control sample according to its statistical uncertainty. We take the width of the variations of fit yields from the nominal fit divided by fit errors on the signal yields and propagate it on the branching fraction.
We take the statistical uncertainty on the expected nonresonant yields determined with a fit to the $\omega$ invariant mass sidebands as a systematic uncertainty. 

To test the modeling of the background exponential \pdf, we perform fits using independent subsets composed of background simulated data and signal generated from the signal $F_{\rm CB}$ \pdf.
An average bias on the signal yield is obtained in the subsets. 
We take the uncertainty on the average bias divided by the signal yield as a systematic uncertainty due to imperfect background modeling and propagate it on the branching fraction.

We propagate the uncertainty on the branching fractions of the \jpsill, ${\omega\to\pip\pim\piz}$, and ${\piz\to\gamma\gamma}$ subdecays used to reconstruct the signal~\cite{ParticleDataGroup:2024cfk}.
The uncertainty on the number of \Bz mesons in the sample comes from the determination of the number of \BBbar pairs and from the knowledge of the $\fpm/\fzz$ ratio~\cite{banerjee2024averagesbhadronchadrontaulepton}.

\begin{table}
\caption{Relative systematic uncertainties on the branching fraction $\BF(\jpsiomega)$ compared with the statistical uncertainty.}
\label{tab:bf-sys}
\begin{tabular*}{\linewidth}{@{\extracolsep{\fill}}lrr}\hline
\hline
Source & \multicolumn{1}{c}{Relative uncertainty $[\%]$} \\
\hline
Tracking efficiency & 1.1\\
\piz reconstruction & 3.8\\
Lepton identification & $\substack{+0.7 \\ -0.5}$\\
\jpsi vertex fit & 0.2\\
$\omega$ vertex fit & 1.2\\
$\omega$ mass window & 1.5\\
BDT selection & 1.2\\
Longitudinal polarization fraction & $\substack{+1.1 \\ -1.0}$\\
Signal model & 0.6\\
Fixed \pdf parameters & 3.0\\
Nonresonant background & 2.1 \\
Background model & 1.4\\
External inputs & 0.9\\
$\numbb$ & 1.5\\
$\fpm/\fzz$ & 1.5\\
\hline
Total systematic uncertainty & 6.6\\
\hline
Statistical uncertainty & 14.0\\
\hline
\hline
\end{tabular*}
\end{table}

\section{Conclusion}
To summarize, we have measured the branching fraction of \jpsiomega decays using data from the \belletwo experiment.
We find \njpsiom signal candidates in a sample of \nbb \BBbar events, corresponding to a decay branching fraction of
\begin{equation}
\bf(\jpsiomega)=\left( \bfval \pm \bferr \pm \bfsys \right) \times 10^{\bfpower}
\end{equation}
where the first uncertainty is statistical and the second is systematic.
The total signal significance is $\sensitivitysysnosigma$ standard deviations, constituting the first observation of the decay.
The significance is calculated by adding the statistical and systematic uncertainties in quadrature.
The results are the most precise to date and consistent with earlier determinations by \lhcb~\cite{LHCb:2012cw,PhysRevD.90.012003}.
They open up the possibility of conducting a \CP violation study with more data from \belletwo.

\section*{Acknowledgements}

% Check with Tom the list of acknowledgments is up to date
% Policy from October 20, 2022
This work, based on data collected using the Belle II detector, which was built and commissioned prior to March 2019,
%Belle1 and data collected using the Belle detector, which was operated until June 2010,
was supported by
%Armenia
Higher Education and Science Committee of the Republic of Armenia Grant No.~23LCG-1C011;
%Australia
Australian Research Council and Research Grants
No.~DP200101792, % Jackson
No.~DP210101900, % Urquijo
No.~DP210102831, % Sevior
No.~DE220100462, % Hsu
No.~LE210100098, % Infrastructure
and
No.~LE230100085; % Infrastructure
%Austria
Austrian Federal Ministry of Education, Science and Research,
Austrian Science Fund
No.~P~34529,
No.~J~4731,
No.~J~4625,
and
No.~M~3153,
and
Horizon 2020 ERC Starting Grant No.~947006 ``InterLeptons'';
%Canada
Natural Sciences and Engineering Research Council of Canada, Compute Canada and CANARIE;
%China
National Key R\&D Program of China under Contract No.~2022YFA1601903,
National Natural Science Foundation of China and Research Grants
No.~11575017,
No.~11761141009,
No.~11705209,
No.~11975076,
No.~12135005,
No.~12150004,
No.~12161141008,
No.~12475093,
and
No.~12175041,
and Shandong Provincial Natural Science Foundation Project~ZR2022JQ02;
%Czech Republic
the Czech Science Foundation Grant No.~22-18469S 
and
Charles University Grant Agency project No.~246122;
%EU
European Research Council, Seventh Framework PIEF-GA-2013-622527,
Horizon 2020 ERC-Advanced Grants No.~267104 and No.~884719,
Horizon 2020 ERC-Consolidator Grant No.~819127,
Horizon 2020 Marie Sklodowska-Curie Grant Agreement No.~700525 ``NIOBE''
and
No.~101026516,
and
Horizon 2020 Marie Sklodowska-Curie RISE project JENNIFER2 Grant Agreement No.~822070 (European grants);
%France
L'Institut National de Physique Nucl\'{e}aire et de Physique des Particules (IN2P3) du CNRS
and
L'Agence Nationale de la Recherche (ANR) under grant ANR-21-CE31-0009 (France);
%Germany
BMBF, DFG, HGF, MPG, and AvH Foundation (Germany);
%India
Department of Atomic Energy under Project Identification No.~RTI 4002,
Department of Science and Technology,
and
UPES SEED funding programs
No.~UPES/R\&D-SEED-INFRA/17052023/01 and
No.~UPES/R\&D-SOE/20062022/06 (India);
%Israel
Israel Science Foundation Grant No.~2476/17,
U.S.-Israel Binational Science Foundation Grant No.~2016113, and
Israel Ministry of Science Grant No.~3-16543;
%Italy
Istituto Nazionale di Fisica Nucleare and the Research Grants BELLE2,
and
the ICSC – Centro Nazionale di Ricerca in High Performance Computing, Big Data and Quantum Computing, funded by European Union – NextGenerationEU;
%Japan
Japan Society for the Promotion of Science, Grant-in-Aid for Scientific Research Grants
No.~16H03968,
No.~16H03993,
No.~16H06492,
No.~16K05323,
No.~17H01133,
No.~17H05405,
No.~18K03621,
No.~18H03710,
No.~18H05226,
No.~19H00682, % Niigata
No.~20H05850,
No.~20H05858,
No.~22H00144,
No.~22K14056,
No.~22K21347,
No.~23H05433,
No.~26220706,
and
No.~26400255,
%the National Institute of Informatics, and Science Information NETwork 5 (SINET5), 
and
the Ministry of Education, Culture, Sports, Science, and Technology (MEXT) of Japan;  
%Korea
National Research Foundation (NRF) of Korea Grants
No.~2016R1-D1A1B-02012900,
No.~2018R1-A6A1A-06024970,
No.~2021R1-A6A1A-03043957,
No.~2021R1-F1A-1060423,
No.~2021R1-F1A-1064008,
No.~2022R1-A2C-1003993,
No.~2022R1-A2C-1092335,
No.~RS-2023-00208693,
No.~RS-2024-00354342
and
No.~RS-2022-00197659,
Radiation Science Research Institute,
Foreign Large-Size Research Facility Application Supporting project,
the Global Science Experimental Data Hub Center, the Korea Institute of
Science and Technology Information (K24L2M1C4)
and
KREONET/GLORIAD;
%Malaysia
Universiti Malaya RU grant, Akademi Sains Malaysia, and Ministry of Education Malaysia;
%Mexico
% CINVESTAV-IPN, UNAM, UAS, BUAP and CONACYT are funded under
Frontiers of Science Program Contracts
No.~FOINS-296,
No.~CB-221329,
No.~CB-236394,
No.~CB-254409,
and
No.~CB-180023, and SEP-CINVESTAV Research Grant No.~237 (Mexico);
%Poland
the Polish Ministry of Science and Higher Education and the National Science Center;
%Russia
the Ministry of Science and Higher Education of the Russian Federation
and
the HSE University Basic Research Program, Moscow;
%Saudi Arabia
University of Tabuk Research Grants
No.~S-0256-1438 and No.~S-0280-1439 (Saudi Arabia), and
King Saud University,Riyadh, Researchers Supporting Project number (RSPD2024R873)  
(Saudi Arabia);
%Slovenia
Slovenian Research Agency and Research Grants
No.~J1-9124
and
No.~P1-0135;
%Spain
%Belle1 Ikerbasque, Basque Foundation for Science,
%Belle1 the State Agency for Research of the Spanish Ministry of Science and Innovation through Grant No. PID2022-136510NB-C33,
Agencia Estatal de Investigacion, Spain
Grant No.~RYC2020-029875-I
and
Generalitat Valenciana, Spain
Grant No.~CIDEGENT/2018/020;
%Swiss (Belle 1)
%Belle1 the Swiss National Science Foundation;
%Sweden
The Knut and Alice Wallenberg Foundation (Sweden), Contracts No.~2021.0174 and No.~2021.0299;
%Taiwan
National Science and Technology Council,
and
Ministry of Education (Taiwan);
%Thailand
Thailand Center of Excellence in Physics;
%Turkey
TUBITAK ULAKBIM (Turkey);
%Ukraine
National Research Foundation of Ukraine, Project No.~2020.02/0257,
and
Ministry of Education and Science of Ukraine;
%USA
the U.S. National Science Foundation and Research Grants
No.~PHY-1913789 % Indiana CEEM
and
No.~PHY-2111604, % Luther
and the U.S. Department of Energy and Research Awards
No.~DE-AC06-76RLO1830, % PNNL
No.~DE-SC0007983, % Wayne State
No.~DE-SC0009824, % Florida
No.~DE-SC0009973, % VPI
No.~DE-SC0010007, % Duke
No.~DE-SC0010073, % South Carolina
No.~DE-SC0010118, % Carnegie Mellon
No.~DE-SC0010504, % Hawaii
No.~DE-SC0011784, % Cincinnati
No.~DE-SC0012704, % BNL
No.~DE-SC0019230, % Duke
No.~DE-SC0021274, % Mississippi
No.~DE-SC0021616, % Mississippi
No.~DE-SC0022350, % Louisville
No.~DE-SC0023470; % South Alabama
%last group
and
%Vietnam
the Vietnam Academy of Science and Technology (VAST) under Grants
No.~NVCC.05.12/22-23
and
No.~DL0000.02/24-25.

% Policy from October 20, 2022
These acknowledgements are not to be interpreted as an endorsement of any statement made
by any of our institutes, funding agencies, governments, or their representatives.

We thank the SuperKEKB team for delivering high-luminosity collisions;
the KEK cryogenics group for the efficient operation of the detector solenoid magnet and IBBelle on site;
the KEK Computer Research Center for on-site computing support; the NII for SINET6 network support;
and the raw-data centers hosted by BNL, DESY, GridKa, IN2P3, INFN, 
%Belle1 PNNL/EMSL, 
and the University of Victoria.

\bibliographystyle{apsrev4-1}
\bibliography{references}

%\newpage
%\appendix
%\input{material}

\end{document}